# A scanning SQUID microscope with 200 MHz bandwidth

**Vladimir V Talanov[1], Nesco M Lettsome, Jr.[1], Valery Borzenets[2], Nicolas Gagliolo[1], Alfred B Cawthorne[3] and Antonio Orozco[1]**

[1]Neocera, LLC, Beltsville, MD 20705, USA
[2]SLAC National Accelerator Laboratory, Menlo Park, CA 94025, USA
[3]Trevecca Nazarene University, Nashville, TN 37210, USA

E-mail: talanov@neocera.com

**Abstract**. We developed a scanning DC SQUID microscope with novel readout electronics capable of wideband sensing RF magnetic fields from 50 to 200 MHz and simultaneously providing closed-loop response at kHz frequencies. To overcome the 20 MHz bandwidth limitation of traditional closed-loop SQUIDs, a flux-modulated closed loop simultaneously locks the SQUID quasi-static flux and flux-biases the SQUID for amplification of the RF flux up to $\Phi_0/4$ in amplitude. Demodulating the SQUID voltage with a double lock-in technique yields a signal representative of both the amplitude and phase of the RF flux. This provides 80 dB of a linear dynamic range with the flux noise density of 4 $\mu\Phi_0/\sqrt{Hz}$ at 200 MHz for $YBa_2Cu_3O_7$ bi-crystal SQUID at 77 K. We describe the electronics performance and present images for RF magnetic field of the travelling wave in a coplanar waveguide, the standing wave in an open-circuited microstrip, and a surface mounted device antenna.

85.25.Dq, 85.25.-j, 68.37.-d, 84.40.Az, 84.40.Ba, 07.20.Mc, 07.55.Ge

## 1. Introduction

DC SQUIDs have found prominent application as ultra-sensitive detectors of magnetic field in a variety of fields, including biology, medicine, metrology, magnetic resonance imaging, and scanning SQUID microscopy (see [1] and references therein). In the resistive state the SQUID voltage is a periodic function of magnetic flux, with a period equal the magnetic flux quantum $\Phi_0$:

$$V = I_c R \sqrt{(I_b/2I_c)^2 - \cos^2(\pi\Phi/\Phi_0)} \quad (1)$$

where $I_c$ and $R$ are the Josephson junction critical current and normal resistance, respectively, and $I_b$ is the SQUID biasing current. In order to linearize the transfer function (1) and improve the sensitivity, practical SQUIDs utilize an external closed loop (flux-locked loop) [2]. Here, magnetic flux typically oscillating at kHz frequency is applied to the SQUID via a modulation coil. The SQUID output is amplified, demodulated by lock-in detector, integrated, inverted, and converted into a current which is fed back into the modulation coil. If the SQUID quasi-static flux equals $n\Phi_0$, $n = 0,1,2,...$, the lock-in output is zero since the SQUID voltage contains no fundamental harmonic. If the quasi-static flux is greater or less than $n\Phi_0$, a fundamental harmonic present in the SQUID voltage makes the lock-in output positive or negative, respectively. Thus, the feedback signal produces a compensating flux equal and opposite to the quasi-static flux offset from $n\Phi_0$, which serves as a measure of the magnetic field.

Although DC SQUIDs possess GHz intrinsic bandwidth (exploited, for instance, by microwave SQUID amplifiers [3]), the transmission line delay between the SQUID sensor and room temperature readout electronics fundamentally limits the closed-loop bandwidth to around 20 MHz [2]. While substantial efforts had been spent increasing this up to 100 MHz [4], it was finally concluded that wideband linearization of the closed-loop SQUIDs is not feasible [5]. Conversely, several scanning SQUID microscopes have been established for imaging microwave magnetic fields in the open-loop regime. R. Black, *et al.* made use of the SQUID nonlinearity to rectify magnetic fields up to 50 MHz and imaged their amplitude [6]. The same group also developed a scanning SQUID microwave microscope in which the frequency is set by the Josephson relation and is continuously tunable up to 200 GHz [7]. J. Matthews, *et al.* designed a scanning SQUID microscope capable of imaging GHz magnetic fields by using a hysteretic SQUID with the pulsed sampling technique [8]. Nevertheless, the open-loop nature of these techniques impedes their practical utility due to limited linearity response and/or susceptibility to variations in the background (static) magnetic field.

Another problem commonly associated with SQUID microwave applications is a parasitic coupling of electromagnetic field to the SQUID wiring [3, 9], which exacerbates at RF frequencies where the size of entire system (*e.g.*, cryogenic dip probe) is comparable to the radiation wavelength (1 m at 300 MHz). Also, the impedance mismatch between the SQUID and electronics may affect the system gain and noise as well. An additional impetus for our work has been given by advances in the near-field microwave microscopy [10], where despite a broad variety of near-field probe designs [11] only few microscopes have been associated with *magnetic* imaging, thus far utilizing a passive loop probe [12] which sensitivity and spatial resolution are much inferior to that of the SQUID.

Our paper is concerned with the development of a scanning SQUID microscope capable of imaging RF magnetic fields in unshielded environment with performance of a closed-loop system. The paper is organized as follows. In the next chapter we will reveal a principle of sensing the RF flux with DC SQUID. Then, we will describe design of readout electronics and scanning SQUID microscope, and evaluate the electronics performance. Finally, we will demonstrate images validating the microscope capabilities in failure analysis of integrated circuits (ICs) and design verification of RF ICs.

## 2. Principle of detecting RF flux with closed-loop DC SQUID

Consider an optimally biased DC SQUID in which the quasi-static flux is "locked" at $n\Phi_0$, $n = 0,1,2,...$ by a flux-modulated closed loop with the modulation flux $\Phi_m \sin(\omega_m t)$ (see figure 1). Since the loop cannot respond to signals outside its bandwidth, the application of RF magnetic flux $\Phi_{RF} \sin(\omega_{RF} t + \varphi_{RF})$ with frequency $\omega_{RF} \gg \omega_m$ makes the net SQUID flux $\Phi_{RF} \sin(\omega_{RF} t + \varphi_{RF}) + \Phi_m \sin(\omega_m t) + n\Phi_0$. If $\Phi_m \sim \Phi_0/4$ the modulating signal flux-biases the SQUID at maximum slopes, positive or negative, of transfer function (1). Then, for $\Phi_{RF} < \Phi_0/4$ the SQUID produces a binary-phase modulated voltage (between 0° and 180°) with the carrier and modulation frequencies of $\omega_{RF}$ and $\omega_m$, respectively [13]. Readout electronics shown in figure 2 demultiplexes the SQUID voltage with a bias-T, which DC output, representative of the SQUID quasi-static flux, is fed into the closed loop. The bias-T RF output is demodulated by an RF lock-in amplifier that feeds an intermediate frequency (IF) lock-in referenced to the modulation frequency. As shown in Appendix, the output of IF lock-in, to be called an IF signal, is proportional to the amplitude and phase of the RF magnetic flux as

$$V_{IF} = G\Phi_{RF} \cos\varphi_{RF} \qquad (2)$$

where $G$ is the small-signal gain of the entire magnetometer, and $\varphi_{RF}$ is the RF flux phase. Setting $\varphi_{RF} = 0$ maximizes the IF signal.

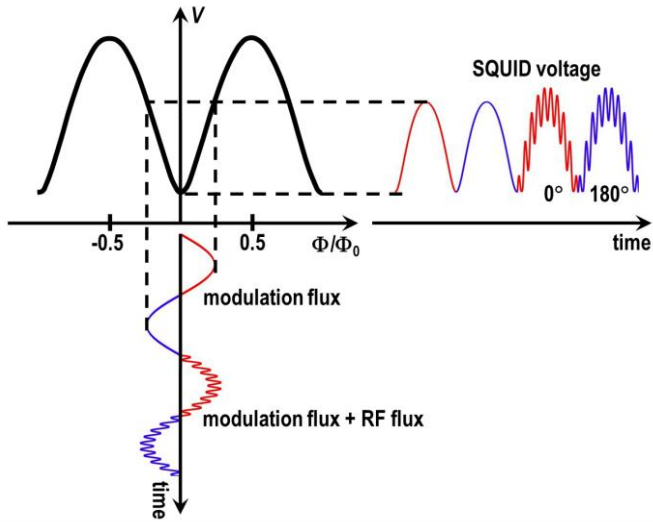

**Figure 1.** Principle of detecting RF flux with DC SQUID. For clarity, the first period of a modulation flux and corresponding output voltage is shown without the respective RF flux and RF voltage superimposed. The quasi-static flux is zero. The output RF voltage carries phase modulation between 0° and 180°.

### 3. Experimental

*3.1. SQUID and electronics*

A direct-coupled $YBa_2Cu_3O_7$ SQUID on bi-crystal $SrTiO_3$ substrate with effective loop area of 30×30 μm$^2$ and single modulation coil has the gain and flux noise density of 20 μV/$\Phi_0$ and <15 μ$\Phi_0$/√Hz at 10 kHz, respectively, at 77K. The SQUID chip is glued onto a bullet-shaped sapphire rod and the SQUID and modulation coil are wire-bonded to differential transmission lines made of cryogenic coaxial cable. Since the SQUID dynamic resistance (1-2 Ohms) is much less than the coax characteristic impedance (50 Ω), the differential connection provides broadband impedance matching between the SQUID and RF electronics.

Readout electronics shown in figure 2 involves a flux-modulated closed loop utilizing 2 kHz modulation frequency and a double lock-in portion associated with demodulation of RF signal. A pair of bias-Ts demultiplexes the SQUID voltage into DC (low-frequency) and RF signals. The bias-Ts DC output is fed into the closed loop, in which the integrator signal is proportional to the SQUID quasi-static flux. After pre-amplification with a low-noise amplifier (LNA), the RF signal is demodulated by an RF lock-in amplifier, which output is fed into an IF lock-in amplifier.

*3.2. Scanning SQUID microscope*

The SQUID sensor and electronics were installed into a commercial scanning SQUID microscope platform described in detail elsewhere [14]. The SQUID RF assembly was cooled down to 77 K by a closed-cycle Joule-Thomson refrigerator and mounted inside a vacuum housing near the 25-μm-thick diamond window, which provides thermal isolation of the SQUID while allows samples to be imaged in air at room temperature (see figure 3). The samples are mounted onto a scanning *xy*-table with 300×150 mm travel and can be brought to within about 50 μm of the window for imaging the component of magnetic field normal to the sample, $B_z$. An inversion algorithm based on a spatial Fourier transform of the Biot-Savart law may be employed to convert the magnetic field image into a 2D current density map [15].

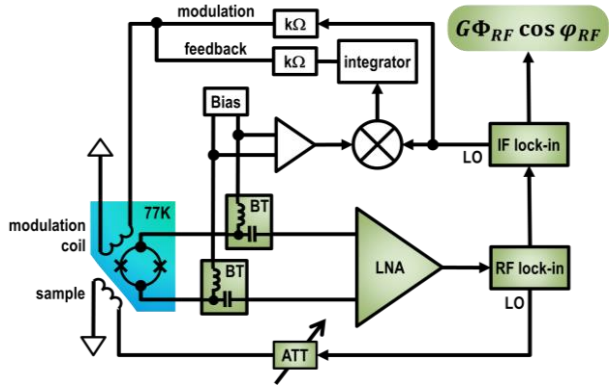
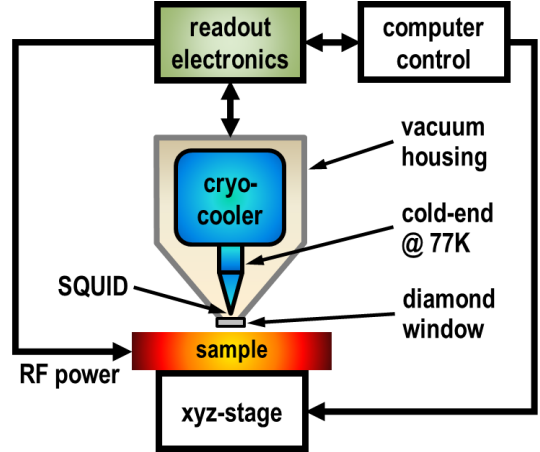

**Figure 2.** Schematic of readout electronics. The shaded parts are associated with processing the RF signal. BT is the bias-T, LNA is the low-noise amplifier, ATT is the variable attenuator, LO is the lock-in local oscillator.

**Figure 3.** Schematic of the scanning SQUID microscope with a room-temperature sample. The SQUID loop is oriented parallel to the sample, to image the magnetic field component $B_z$ normal to a sample.

*3.3. Method verification*

To verify that the IF signal is due to RF flux threading the SQUID, rather than a parasitic pick-up caused by inductive coupling of RF magnetic field to the SQUID wiring, we acquired the IF lock-in output vs. quasi-static flux with the closed-loop *open* (see figure 4). The quasi-static flux was produced by variable DC current applied to the SQUID modulation coil. The RF flux of known amplitude was produced by feeding RF signal into the modulation coil (rather than a sample) via a variable attenuator. Approximating the coil as a short terminating a feedline yields $\Phi_{RF} = g2\sqrt{P_{RF}/Z_0}$, where $g$ is the geometrical coefficient relating the coil current to the SQUID flux, $P_{RF}$ is the incident RF power, and $Z_0$ is the feedline characteristic impedance. Setting $\Phi_m, \Phi_{RF} \ll \Phi_0$ and $I_b = 2I_c$ provides a dependence representative of the second derivative of the SQUID transfer function, which is $\partial^2 V/\partial \Phi^2 = 2\pi^2 I_c R \cos[2\pi\Phi/\Phi_0]/\Phi_0^2$ from (A.1). No offset about zero in the IF signal confirms that a double lock-in demodulation efficiently removes the parasitic RF signal.

In order to characterize the sensitivity and linearity of our method, the IF signal was measured vs. RF flux at 165 MHz with the closed-loop *closed*, which was "locking" the SQUID quasi-static flux at $n\Phi_0$ (see figure 5). The bias current, modulation flux, and RF phase were optimized at $I_b = 2I_c$, $\Phi_m \approx \Phi_0/4$ and $\varphi_{RF} = 0$, respectively. At smaller RF flux, the dependence is linear, limited by the flux noise density of $4\times10^{-6}$ $\Phi_0/\sqrt{Hz}$, which is close to the theoretical white noise limit $\sim 10^{-6}\Phi_0/\sqrt{Hz}$ for our SQUID. At larger RF flux, the SQUID non-linearity causes a maximum at $\Phi_{RF} \simeq 0.2\Phi_0$, which agrees with that of $0.21\Phi_0$ predicted by (A.5). Fitting experimental data to the following equation, obtained from (A.5),

$$V_{IF} = G\Phi_{RF}\left(1 - \frac{3\pi^2}{4}\frac{\Phi_{RF}^2}{\Phi_0^2}\right) \quad (3)$$

yields the small signal gain for the entire magnetometer of $G = 62$ V/$\Phi_0$ (see figure 5, inset). Thus, our technique provides four orders (80 dB) of usable dynamic range, from $10^{-5}\Phi_0$ to $0.1\Phi_0$, with 7% non-linearity. This is sufficient for most applications associated with the imaging of RF magnetic field, as will be demonstrated in the next section.

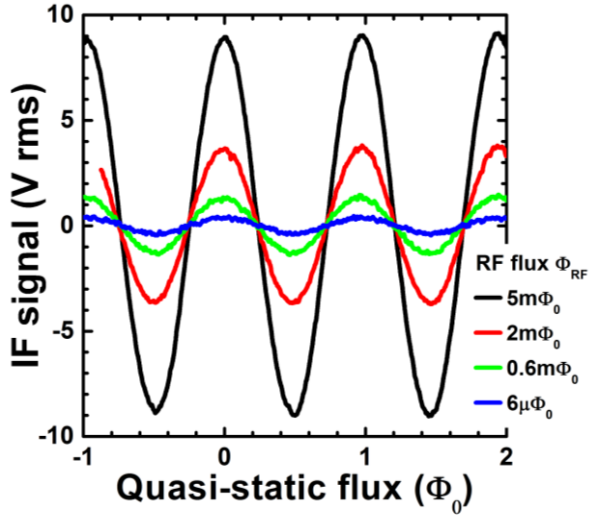
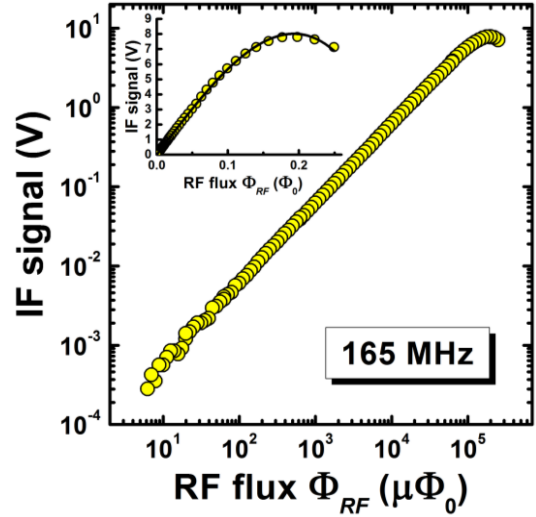

**Figure 4.** IF signal versus SQUID's quasi-static flux with the closed loop held open, for several amplitudes of RF flux, $\Phi_{RF}$, at 190 MHz.

**Figure 5.** Log-log plot of the apparatus transfer function at 165 MHz, 77 K. Inset: the same data on a linear scale; solid line is a fit to (3).

## 4. Imaging examples

### 4.1. Travelling wave in CPW

Figure 6 shows the result of imaging the *z*-component of magnetic field in a coplanar waveguide (CPW). The waveguide was fed with 1 mW of power at 200 MHz from one end, while the other end was terminated with a matched load (see cartoon inset), in order to allow only a propagating wave to exist in the sample. Since the sample size was much less than the radiation wavelength, only a portion of the wave appears in the image. The RF phase $\varphi_{RF}$ was adjusted such that the wave crest is present within the scanning area.

### 4.2. Standing wave in open-circuited microstrip

Figure 7 shows the result of imaging the *z*-component of magnetic field in an open-circuited microstrip transmission line fed with 1 mW at 190 MHz. The open boundary was created by a 50 μm gap cut across entire 1.3-mm-wide microstrip. Since the gap impedance is virtually infinite, the open load reflects entire incident power back toward the generator such that the reflected current wave is 180 degrees out of phase with the incident current wave. This creates a full standing wave, in which the current amplitude varies vs. position *x* along the microstrip varies as $I = 2\sqrt{P/Z_0}\sin(\gamma x)$, with *P* the incident RF power, $Z_0$ the microstrip characteristic impedance, and *γ* the microstrip propagation constant [16]. Since the sample was much smaller than the radiation wavelength $\lambda = 2\pi/\gamma$, within the image the current varies linearly due to $\sin(\gamma x) \approx \gamma x$. The current density map in figure 6 was obtained from the magnetic field using the inversion algorithm [15]. The first node (valley) in the current standing wave pattern coincides with the gap. This phenomenon has been employed elsewhere for failure analysis (FA) of open defects in logic and memory ICs [17, 18].

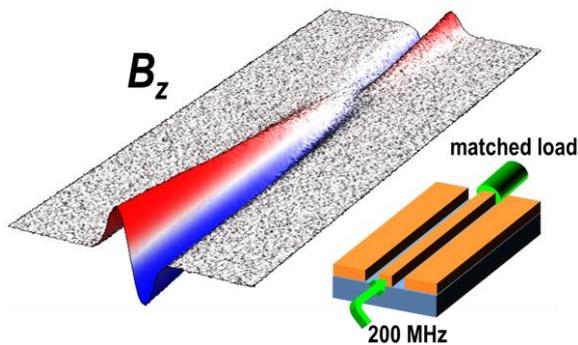 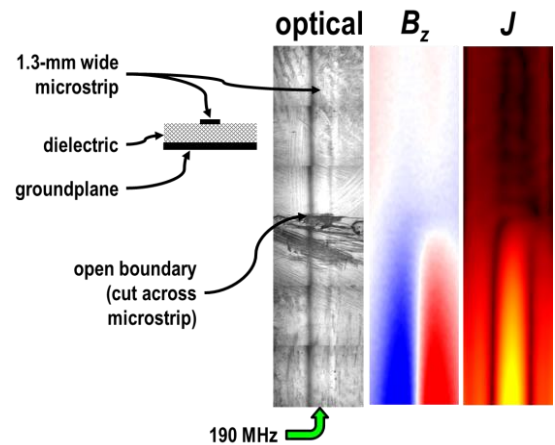

**Figure 6.** A 3D intensity map for the *z*-component of magnetic field in a travelling wave in CPW at 200 MHz. The red and blue regions represent the "up" and "down" direction of the magnetic field.

**Figure 7.** A standing wave in the open-circuited microstrip at 190 MHz. Left to right: microstrip cross-section; microstrip optical image; 2D intensity map for the *z*-component of magnetic field $B_z$; current density image obtained by inversion.

*4.3. Electrically small SMD antenna*

Figure 8 shows the results of imaging a surface mounted device (SMD) antenna with nominal bandwidth of 164–175 MHz [19] installed onto an evaluation board. The 25-mm-long antenna was fed with 1 mW of power at 169 MHz via CPW with impedance matching circuit. The current density map was obtained from the magnetic field using the inversion algorithm [15]. A groundplane that was incidentally formed by an aluminum chuck holding the sample shifted the actual antenna bandwidth away from its nominal range, which reduced the antenna return loss from the specification of 12 dB down to less than 1 dB, causing an anomalous displacement current through one of the capacitors in the impedance matching circuit.

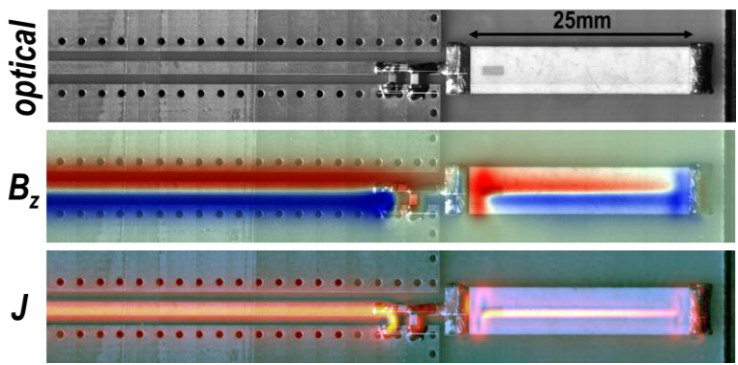

**Figure 8.** Commercial SMD antenna at 169 MHz. Top to bottom: an optical image of the antenna with impedance matching circuit fed by CPW; intensity map for the *z*-component of magnetic field $B_z$ overlaid with the optical image; current density image overlaid with the optical image.

**5. Conclusion**

We designed a scanning DC SQUID RF microscope capable of imaging alternating magnetic fields from 50 to 200 MHz. The lower frequency was limited by the bandwidth of our RF pre-amplifier and may be brought into 100s kHz range, provided the modulation and RF frequencies are spread far enough for demultiplexing. The upper frequency was limited by the bandwidth of the commercial RF lock-in and may potentially be extended into GHz range. A direct-coupled 30-µm-sized DC SQUID at 77 K affords the noise density of 8 pT/√Hz at 200 MHz. A flux-modulating closed loop compensates for variations in the background magnetic field, which allows operating the SQUID in unshielded environment. A linear dynamic range of four orders in magnitude enables such applications as failure analysis of semiconductor ICs and near-field imaging of RF SMDs.

**Acknowledgments**
Authors acknowledge E. Wong and S. Garrahan for assistance with the experimental setup, and J. Matthews, Prof. T. Venkatesan and J. Gaudestad for valuable discussions. V.V.T. is indebted to A. Stepanova and A. Grafov for their hospitality. This work has been supported by NSF-SBIR IIP-0924610.

**Appendix. Theory of a double lock-in demodulation**
The SQUID transfer function (1) can be approximated by

$$V = \frac{RI_c}{4}\left[\frac{I_b}{I_c} + \sqrt{\frac{I_b^2}{I_c^2} - 4} + \left(\sqrt{\frac{I_b^2}{I_c^2} - 4} - \frac{I_b}{I_c}\right)\cos\frac{2\pi\Phi}{\Phi_0}\right] \tag{A.1}$$

Inserting $I_b = 2I_c$ into (A1) and expanding the result around $\Phi = 0$ yields

$$V \simeq G_{SQUID}\Phi_0\pi\left[\left(\frac{\Phi}{\Phi_0}\right)^2 - \frac{\pi^3}{3}\left(\frac{\Phi}{\Phi_0}\right)^4\right] \tag{A.2}$$

where $G_{SQUID} = (\partial V/\partial \Phi)_{\Phi_0/4} = \pi RI_c/\Phi_0$ is the SQUID gain at $\Phi = \Phi_0/4$. The net SQUID flux is a sum of the modulation flux $\Phi_m\sin\omega_m t$ and the RF flux $\Phi_{RF}\sin(\omega_{RF}t + \varphi_{RF})$. Inserting this into (A.2), multiplying the result by the LNA voltage gain $G_{LNA}$ and RF lock-in reference $G_{RF}\sin\omega_{RF}t$, and retaining only DC and low frequency terms yields for the RF lock-in output:

$$V_{RF} = G_{SQUID}G_{LNA}G_{RF}\pi\left[1 - \frac{\pi^2\Phi_{RF}^2}{2\Phi_0^2} + \frac{2\pi^2\Phi_m^2}{3\Phi_0^2}\cos^2(\omega_m t)\right]\frac{\Phi_m}{\Phi_0}\sin(\omega_m t)\Phi_{RF}\cos\varphi_{RF} \tag{A.3}$$

where $G_{RF}$ is the RF lock-in gain. Multiplying (A.3) by the IF lock-in reference $G_{IF}\sin\omega_m t$ and retaining only DC terms yields for the IF lock-in output:

$$V_{IF} = G_{SQUID}G_{LNA}G_{RF}G_{IF}\frac{\pi}{2}\left[1 - \frac{\pi^2(\Phi_m^2 + \Phi_{RF}^2)}{2\Phi_0^2}\right]\frac{\Phi_m}{\Phi_0}\Phi_{RF}\cos\varphi_{RF} \tag{A.4}$$

where $G_{IF}$ is the IF lock-in gain. Using $\Phi_m = \sqrt{2}\Phi_0/\sqrt{3}\pi \approx 0.26\Phi_0$ maximizes the IF signal:

$$V_{IF} = \frac{2}{3\sqrt{6}}G_{SQUID}G_{LNA}G_{RF}G_{IF}\left(1 - \frac{3\pi^2}{4}\frac{\Phi_{RF}^2}{\Phi_0^2}\right)\Phi_{RF}\cos\varphi_{RF} \tag{A.5}$$

For small RF fluxes with amplitude $\Phi_{RF} \ll \Phi_0$ we finally obtain from (A.5):

$$V_{IF} = G\Phi_{RF}\cos\varphi_{RF} \tag{A.6}$$

where $G = 0.27 G_{SQUID}G_{LNA}G_{RF}G_{IF}$ is the small signal gain of entire apparatus.

Accepted for publication in Superconductor Science and Technology, 2014**References**
[1]   Clarke J and Braginski A 2004 *The SQUID Handbook: Fundamentals and Technology of SQUIDs and SQUID Systems* (Weinheim: Wiley-VCH Verlag GmbH & Co. KGaA)
[2]   Drung D 2003 High-$T_c$ and low-$T_c$ dc SQUID electronics *Supercond. Sci. Technol.* **16** 1320
[3]   Mück M, Christian W C and Clarke J 2003 Superconducting quantum interference device amplifiers at gigahertz frequencies *Appl. Phys. Lett.* **82** 3266
[4]   Drung D, Aßmann C, Beyer J, Peters M, Ruede F and Schurig Th 2005 dc SQUID readout electronics with up to 100 MHz closed-loop bandwidth *IEEE Trans. Appl. Supercond.* **15** 777
[5]   Kornev V K, Soloviev I I, Klenov N V and Mukhanov O A 2009 Bi-SQUID: a novel linearization method for dc SQUID voltage response *Supercond. Sci. Technol.* **22** 114011
[6]   Black R C, Wellstood F C, Dantsker E, Miklich A H, Koelle D, Ludwig F and Clarke J 1995 Imaging radio-frequency fields using a scanning SQUID microscope *Appl. Phys. Lett.* **66** 1267
[7]   Black R C, Wellstood F C, Dantsker E, Miklich A H, Nemeth D T, Koelle D, Ludwig F and Clarke J 1995 Microwave microscopy using a superconducting quantum interference device *Appl. Phys. Lett.* **66** 99
[8]   Matthews J, Vlahacos C P, Kwon S P and Wellstood F C 2005 Sampling method to extend bandwidth of scanning SQUID microscopes *IEEE Trans Appl Supercond* **15** 688; Vlahacos C P, Matthews J and Wellstood F C 2011 A Cryo-cooled scanning SQUID microscope for imaging high-frequency magnetic fields *IEEE Trans. Appl. Supercond.* **21** 412
[9]   Ranzani L, Spietz L and Aumentado J 2013 Broadband calibrated scattering parameters characterization of a superconducting quantum interference device amplifier *Appl. Phys. Lett.* **103** 022601
[10]  Anlage S M, Talanov V V and Schwartz A R 2007 Principles of near-field microwave microscopy *Scanning Probe Microscopy: Electrical and Electromechanical Phenomena at the Nanoscale* vol. 1, ed S Kalinin and A Gruverman (Springer Science, New York, 2007) pp. 215-252
[11]  Rosner B T and van der Weide D W 2002 High-frequency near-field microscopy *Rev. Sci. Instrum.* **73** 2505
[12]  Lee S-C, Vlahacos C P, Feenstra B J, Schwartz A R, Steinhauer D E, Wellstood F C and Anlage S M 2000 Magnetic permeability imaging of metals with a scanning near-field microwave microscope *Appl. Phys. Lett.* **77** 4404
[13]  Talanov V, Lettsome N, Orozco A, Cawthorne A and Borzenets V 2012 DC SQUID RF magnetometer with 200 MHz bandwidth *American Physical Society, APS March Meeting (Boston, MA, USA, February 27-March 2, 2012)*, abstract #Q54.003
[14]  Knauss L A, Orozco A, Woods S I and Cawthorne A B 2003 Advances in scanning SQUID microscopy for die-level and package-level fault isolation *Microelectronics Reliability* **43** 1657
[15]  Roth B J, Sepulveda N G and Wikswo J P 1989 Using a magnetometer to image a two-dimensional current distribution *J. Appl. Phys.* **65** 361; Chatraphorn S, Fleet E F, Wellstood F C, Knauss L A and Eiles T M 2000 Scanning SQUID microscopy of integrated circuits *Appl. Phys. Lett.* **76** 2304
[16]  Pozar D M *Microwave Engineering* (John Wiley & Sons, Inc, 2nd ed, 1998, NY) p. 69
[17]  US and worldwide patents are pending
[18]  Gaudestad J, Talanov V V and Huang P C 2012 Space domain reflectometry for opens detection location in microbumps, *Microelectronics Reliability* **52** 2123
[19]  ISM Chip Antenna 164 – 175 MHz, Johanson Technology, Inc.8